# Parallel Spectral Clustering Algorithm Based on Hadoop

**Yang Zhao，Kafei Xiao，Yajun Cui，Lei Wang，Chenglong Zhang**

**University of Chinese Academy of Sciences**

**Beijing，China**

**Abstract：Spectral clustering and cloud computing is emerging branch of computer science or related discipline. It overcome the shortcomings of some traditional clustering algorithm and guarantee the convergence to the optimal solution, thus have to the widespread attention. This article first introduced the parallel spectral clustering algorithm research background and significance, and then to Hadoop the cloud computing Framework has carried on the detailed introduction, then has carried on the related to spectral clustering is introduced, then introduces the spectral clustering arithmetic Method of parallel and relevant steps, finally made the related experiments, and the experiment are summarized.**

**Key words: Spectral Clustering；Cloud Computing；Hadoop；Parallelization**

# Contents

# Chapter 1 Introduction

## 1.1 The Significance of Topic

To understand a new object or a new phenomenon, people always try to find its features, and then compared with the known object or phenomenon, according to certain standards and rules whether similar between them. With the rapid development of Internet, brings us a lot of data and information and at the same time give rise to a large number of data accumulation. In order to better represent and understand these data, using a computer the data classification or effectively. Clustering is particularly important.

Clustering is in accordance with the requirements for a certain similarity to the process of grouping samples. It is a kind of widely used data analysis tools. Comparing with the classification of supervised learning methods, clustering methods have obvious advantage:

1. Collect and tag large sample set is a laborious work and in many cases we can't get data category attributes;

2. For classification samples can change slowly over time, the nature of the change in unsupervised learning cases compared with supervised learning is easier to get, at this time of learning can be able to get a boost;

3. Unsupervised method can extract some of the basic features of these characteristics to the further steps to provide preprocessing and feature extraction of effective early treatment;

4. Clustering algorithm can reveal to us the test data of some internal structure and rules, if we can get some valuable information through these methods, can be more targeted to design better subsequent classifier.

## 1.2 Overseas and Domestic Research Status

In recent years, machine learning algorithm based on graph theory is becoming a new research hotspot in the field of machine learning. Spectral clustering Algorithms can be classified as segmentation method based on graph theory. At present, all kinds of based on graph partition criterion put forward to promote the development of machine learning algorithms. Recently, a clustering method basing on spectral graph theory began to get attention. The class method using similarity matrix feature clustering feature vectors of the decomposed, this kind of algorithm is referred to as spectral clustering. Although the spectral clustering method has achieved good results, but now still in the early stage of development, there are still many algorithm itself many open questions worth further study, and there is no complete theory to explain the spectral clustering algorithm and analyze its innings sex limited. Spectral clustering algorithm although has obtained the good effect, but the algorithm is still in the early stages of the development and the algorithm still has many issues worthy of further research.

## Chapter 2 Introduction of Hadoop

Hadoop is an open mature and widely used open source cloud computing framework. It implements a distributed file system (the Hadoop Distributed File System), hereinafter referred to as the HDFS. It also implements a complete graphs the calculation of distributed programming framework, the framework can use HDFS above for large-scale data analysis and data processing.

### 2.1 HDFS distributed file system

HDFS is a distributed file storage system. Through streaming data access pattern to store large files and large files. Its characteristic is a write, read many times and efficient batch. So the HDFS more focus on high throughput of data access. One of the main design goals of HDFS is under fault condition also can ensure the reliability of data storage. HDFS has a relatively complete redundancy backup and fault recovery mechanism. It be achieved reliably store huge amounts of documents in the clusters.

## 2.2 Parallel computing programming thought graphs

Graphs by adopting the idea of "divide and rule", to the operation of the large-scale data sets, distributed to a master node under the management of each node to complete together, and then through the integration of each node in the middle of the results to get the final result. Highly abstract graphs for two functions: map and reduce, the map is in charge of the task is decomposed into multiple tasks, reduce multitasking is responsible for the decomposition results summary. As for the other total complex problems in parallel programming, such as distributed storage, job scheduling, load balancing and fault tolerance, network communication, etc. It's responsible for handling all by graphs framework.

## 2.3 HBase distributed database

HBase based on HDFS, it can provide high reliability, high performance, columns, storage, scalable, and real-time database system to read and write. It can build from the bottom up, simply by adding nodes to achieve linear scaling. HBase is not a relational database, it does not support SQL. But in specific problem space, it can do a RDBMS cannot do: on cheap hardware cluster management of large scale sparse tables.

## 2.4 Summary

This chapter briefly introduces the two core components of the Hadoop framework. HDFS distributed file system and parallel programming framework graphs as well as build upon HDFS HBase distributed no database. Spectral clustering parallel to the help of the three components of HDFS is used to store the initial input files, graphs are the core of the whole parallel computing system, whereas HBase when matrix is calculated to store intermediate results and the final result.

# Chapter 3 Overview of spectral clustering algorithm

## 3.1 Summary

Spectral clustering algorithm is developed based on clustering algorithm of graph theory recently. Compared with some traditional clustering algorithms, Spectral clustering algorithm has its obvious advantages. It is not only simple, but also effective that can be able to identify the sample space of arbitrary shape. It can converge to the global optimal solution and be well suited to real problems. The reason of success of spectral clustering: Through the similarity matrix or Laplacian matrix Eigen decomposition, we can get a global optimal solution of a clustering criterion in a continuous domain.

## 3.2 Theoretical basis of spectral clustering

The thoughts of spectral clustering is from the spectra divided. Spectral clustering translate clustering problem into multi-channel partitioning problem of undirected graph .Data points are looked as vertices V in undirected graph G (V, E).Weighted edges' set $E = \{S_{ij}\}$ represents the similarity between two vertices based on a similarity metric calculation, which uses S to represent the similarity matrix of clustering data points. In the graph G we translate clustering problem into graphic partitioning problem on the graph G. That's to say, the G (E, V) is partitioned into k disjoint subsets $V_1, V_2, ..., V_k$ to ensure in each subset $V_j$ similarity is high and the similarity between the different sets is low.

### 3.2.1 Basic concepts of graph

Setting up sample points is $\{X_1, X_2, ..., X_n\}$.We define the similarity of between two sample points $X_i$ and $X_j$ is $S_{ij}$.We also use the similarity graph G (V, E) to express the similarity between samples. Each vertex of the graph represents a point $X_i$ .If the weight $S_{ij}$ between the two sample points is greater than 0, it can be considered the connection between the two sample points.

We define G= (V, E) is a undirected weight connected graph. The vertex set is $\{v_1, v_2, ..., v_n\}$ .The connection weight between the vertices $v_i$ and $v_j$ that exists connection is $w_{ij}$ .Weighted adjacency matrix of graph is $W = (w_{ij})(i, j = 1, 2, ..., n)$ .If $w_{ij} = 0$ ,There is no connection between $v_i$ and $v_j$ .If the G is the undirected connection graph, we can reach $w_{ij} = w_{ji}$ .The degree of the vertex is defined as:

$$d_i = \sum_j w_{ij} , \qquad j \in adjacent(i)$$

We define a matrix D that diagonal matrix is consisted by $d_1, d_2, ..., d_n$ .Given a subset of vertices $A \subset V$ .The definition of its complement is $\overline{A}$ .We define an indicator vector $l_A = \{f_1, f_2, ..., f_n\}$ . If $v_i \in A$ , $f_i = 1$ .We also define $i \in A$ which represents $\{i \mid v_i \in A\}$ ,For two subsets $A, B \in V$ ,we define:

$$W(A, B) = \sum_{i \in A, j \in B} w_{ij}$$

Then define two metrics to measure the size of subset $A \subset V$

$$vol(A) = \sum_{i \in A} d_i$$

$vol(A)$ defines the size of the set A by the degree of all vertices in A.

If the set A is connected, any two vertices in the set A can be connected by a path within the set. If there is no connection between the A and the $\overline{A}$ , we define A is a connected component. If $A_i \cap A_j = \phi$ and $A_1 \cup A_2 \cup ... \cup A_k = V$ ,non-empty set $A_1, A_2, ..., A_k$ form a graph partition.

### 3.2.2 Laplasse matrix of graph and its properties

The main tool of spectral clustering is the Laplasse matrix of graph. For the non-normalized Laplasse matrix, we have the following proposition:

We suppose an undirected weight graph G. The number of the eigenvalues of the non-normalized Laplasse matrix L is 0 represents the number of independent components connected for its graph. If it has K eigenvalues whose value is 0, it will have K connected component $A_1, A_2, \ldots, A_k$ and the feature space of feature values is 0 that has a liner combination of Indicator vector $l_{A_1}, l_{A_2}, \ldots, l_{A_k}$ of connected component .

For the normalized Laplasse matrix, we have the following proposition:

We suppose a undirected weight graph G. The number of the eigenvalues of the normalized Laplasse matrix $L_{sym}$ and $L_{rw}$ is 0 represents the number of independent components connected for its graph. If it has K eigenvalues whose value is 0,it will have K connected component $A_1, A_2, \ldots, A_k$ and eigenvalue of $L_{rw}$ is 0 that has a liner combination of Indicator vector $D^{1/2}l_{A_1}, D^{1/2} l_{A2}, \ldots, D^{1/2}l_{A_k}$ of connected component .

### 3.2.3 Realization of spectral clustering algorithm

We use the sample point $x_1, x_2, \ldots, x_n$ to represent sample collection for arbitrary clustering. Similarity is

$$S_{ij} = \exp(-\left\| x_i - x_j \right\|^2) / 2\sigma^2$$

Its corresponding similarity matrix is $S = (s_{ij})(i, j = 1,2, \ldots, n)$.

## Chapter 4 Parallel spectral clustering algorithm

### 4.1 Overview of parallel spectral clustering

After years of continuous research and unremitting exploration, Spectral Clustering has been recognized as a clustering algorithm which is more effective than the traditional clustering algorithm, and its mathematical basis is the graph cut and matrix operation. In general, The time complexity of spectral clustering is O ($n^3$), where n is the number of objects to be entered. Because of its high complexity, it greatly limits its application in the actual production and research.

To reduce the time complexity of spectral clustering, this chapter tries to combine spectral clustering algorithm and MapReduce programming ideas of Hadoop together. Through the analysis of the traditional spectral clustering algorithm steps, we can achieve concurrent steps to separate out and put these steps integration into the MapReduce, combined with Hadoop excellent distributed storage and parallel computing performance, realize the spectral clustering algorithm parallelization, take advantage of the cluster, and reduce the time needed for the clustering ultimately.

## 4.2 Analysis of single spectrum clustering algorithm

According to the similarity measurement, single spectral clustering algorithm can be divided into several different methods. In here, we choose normalized spectral clustering to illustrate.

**Algorithm 4.1 normalized spectral clustering algorithm**

**Input: data points set $x_1$, $x_2$,..., $x_n$, clusters number is k**

**Output：clusters $C_1$, ..., $C_k$**

**1. Calculate the similarity matrix $S \in R^{n \times n}$ , $S(x_i, x_j)$ is data points $x_i$ and $x_j$ similarity and then sparse it**

**2. Constructing diagonal degree matrix D, and diagonal elements are $d_i = \sum_{j=1}^{n} S(x_i, x_j)$**

**3. Calculate normalized Laplasse matrix L, $L = L - D^{-\frac{1}{2}} S D^{-\frac{1}{2}}$**

**4. Calculate k the minimum eigenvectors of L, and the composition matrix $Z \in R^{n \times k}$ contains them.**

**5. Standardized Z to $Y \in R^{n \times k}$**

**6. The data points with K-means algorithm $y_i \in R^k$ (i = 1, 2, ..., n) into k clusters $C_1$,... $C_k$.**

### 4.3 The realization of parallel spectral clustering on Hadoop

### 4.3.1 Parallel computing similarity matrix

**Because Hadoop's MapReduce parallel programming framework can provide excellent distributed computing framework, HBase distributed database building on HDFS can be used to initialize and store intermediate results matrix. So, we choose MapReduce, a core component of the Hadoop, to achieve our parallel spectral cluster with the distributed file system HDFS and HBase distributed database. We first put adjacency matrix which is constitute of the data point $x_1, x_2, \cdots, x_n$ into HBase table, the table can be clustered access to all of the machines, and the key row of each record is set as the index of the data points. Then we use a map function to automatically calculate the similarity between the data points.**

$\forall i, j, 1 \leq i \leq j \leq n,$ **we just need to calculate** $sim(x_i, x_j)$**.**

**Because these objects can constitute undirected graphs--** $sim(x_i, x_j) = sim(x_j, x_i)$**, the calculation of the similarity between each pair of data points needs to be calculated once. And according to the symmetry of undirected graphs, the other half of the similarity values are obtained.**

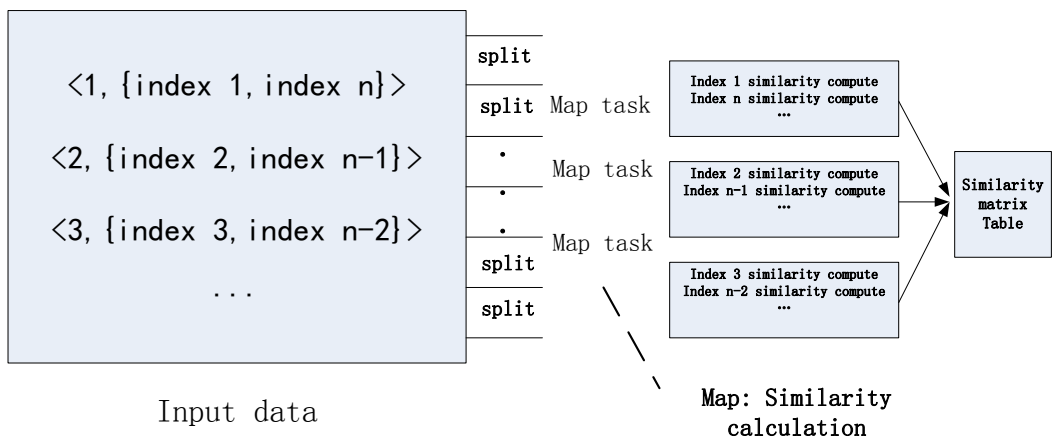

**Fig.1. Map function of Parallel computing similarity matrix**

**Algorithm 4.2 Map function design of constructing similarity matrix**

**Input：<key，value>，key as point index，value as null**

**Output：<key'， value'>=<key, null>**

**1、index = key, anotherindex = n-key+1**

**2、for i in {index, anotherindex}**

    **i_content = getContentFromHBase(i);**

    **for j = i to n do**

        **j_content = getContentFromHBase(j);**

        **sim = computeSimilarity(i_content, j_content);**

        **storeSimilarity(i,j,sim) into HBase table;**

    **End For**

**End For**

**3、Output <key, null>**

**4、End**

    **It should be noted that "similar value of the subscript i" need to calculate the value of $n-i+1$ pairs of data point $\left\{\left\langle x_i, x_i\right\rangle, \left\langle x_i, x_{i+1}\right\rangle, \ldots, \left\langle x_i, x_n\right\rangle\right\}$. Therefore, in order to load balance ,we calculate the similarity of index i and index n-i+1, which performed on the same machine.**

### 4.3.2 Parallel computing k smallest eigenvectors

    **Lanczos algorithm is a kind of algorithm that transforms the symmetry matrix into a symmetric three diagonal matrix by orthogonal transformation, and is named as the Hungarian mathematician Lanczos Cornelius in twentieth Century. Lanczos algorithm is as follows, see algorithm 4.3.**

**Algorithm 4.3 The design idea of Lanczos algorithm**

**1、 $v_1 \leftarrow$ The random vector of norm as 1**

$$v_0 \leftarrow 0$$

$$\beta_1 \leftarrow 0$$

2、 *Iteration* : *for*    $j = 1, 2, \ldots, \text{m}$

$$w_j \leftarrow L v_j - \beta_j v_{j-1}$$

$$\alpha_j \leftarrow \left( w_j, v_j \right)$$

$$w_j \leftarrow w_j - \alpha_j v_j$$

$$\beta_{j+1} \leftarrow \| w_j \|$$

$$v_{j+1} \leftarrow w_j / \beta_{j+1}$$

3、 Re*turn*

Note that $(x, y)$ is the dot product of two vectors, and after iteration, we get a three diagonal matrix of $\alpha_j$ and $\beta_j$ :

$$T_{mm} = \begin{pmatrix} \alpha_1 & \beta_2 & & & & 0 \\ \beta_2 & \alpha_2 & \beta_3 & & & \\ & \beta_3 & \alpha_3 & \cdots & & \\ & & \cdots & \cdots & \beta_{m-1} & \\ & & & \beta_{m-1} & \alpha_{m-1} & \beta_m \\ & & & & \beta_m & \alpha_m \end{pmatrix}$$

After getting the matrix $T_{mm}$, because $T_{mm}$ is a three diagonal matrix, it is easy to get its eigenvalues and eigenvectors by some methods (such as QR). It can be proved that these eigenvalues (eigenvectors) are the similarity values of the eigenvalues (eigenvectors) of the original Laplasse matrix L.

As can be seen from the Lanczos algorithm, $L v_j$, the multiplication of matrix and vector, is a relatively time-consuming process. If the matrix L is not enough to load memory, and each time it is multiplied by a vector to move L, the time consumption will be great.

In the distributed storage and computing framework provided by MapReduce Hadoop and HDFS, an excellent idea is used: "mobile computing is more effective than mobile data". We use a class Matrix Distributed to the matrix L to be decomposed to HBase up, the matrix L on the HBase stored in the time when the line to the segmentation store. Then each iteration of the Lanczos does not to move the HBase distributed storage of the matrix L, instead, moving vector (Mobile Computing), and parallel computing the product of vectors $v_j$ and matrices L by multiplying the rows of matrices L.

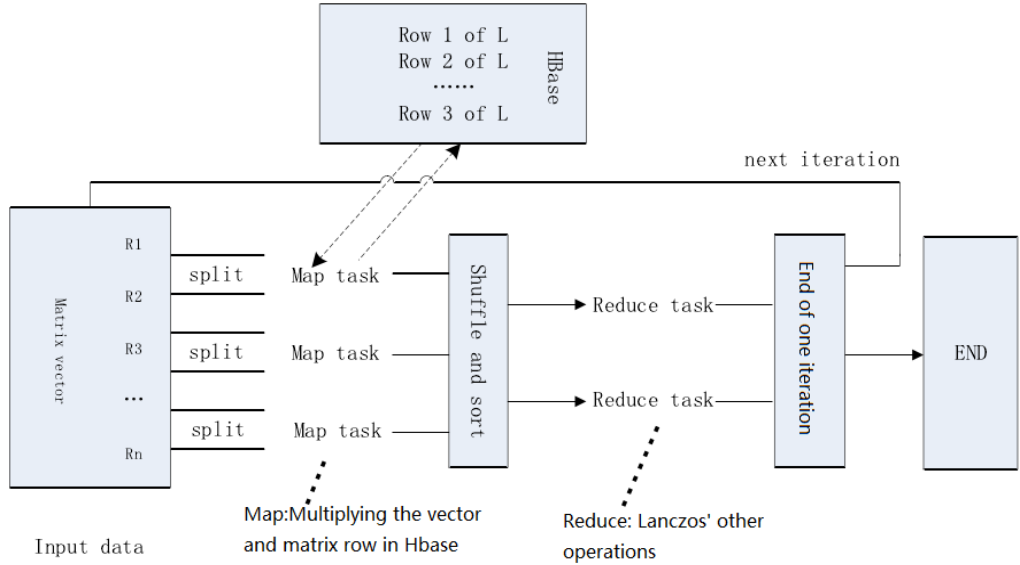

**Fig.2. The Map/Reduce of Parallel computing k smallest eigenvectors**

$Lv_j$ is the main time consuming operation of the Lanczos algorithm, now with the matrix L distributed stored in HBase, this operation can complete in map / reduce function. If the K feature vector is needed, the vector is transferred to the data store of L for K times to compute.

### 4.3.3 Parallelization K-means clustering

Once have gotten the k minimal eigenvector, we can get a new expression form $y_i$ of metadata point $x_i$ by the way of standardization. In the parallel K-means clustering algorithm, a table containing the HBase that initialize the center of the K cluster is created and can be accessed by the individual machines on the HBase. Obviously, a data point and the distance between the k centers are calculated and the other data points and the distance between the K centers are independent of each other. So different data points and k center distance can be executed in parallel computing framework of MapReduce. Based on the Hadoop parallel K-means clustering algorithm design, the main work is the design and implementation of map and reduce functions, including the type of input and output (key, value) and the Map and Reduce functions specific logic etc.

The MapReduce implementation algorithm of brief clustering K-means as the chart below shows:

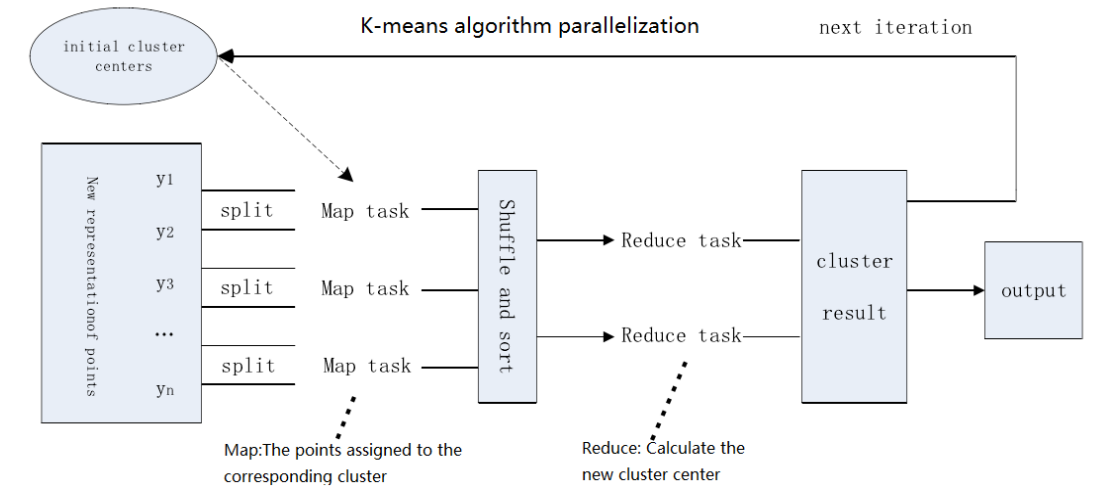

**Fig.3. The clustering K-means algorithm parallelization on Hadoop**

1. Create a file containing the initial cluster center. This file contains the cluster center for each iteration. To the following account conveniently, this file is called "the center file".

2. Map function: read the center file and get the center of the last iteration (or initialization). Read the data points, and calculate the distance between each data point and several centers, then assign each data point to the center of its closest (most similar).

map(<key, value>, <key', value>')

{

from the value to analysis the sample object, recorded as instance;

the maximum value of the auxiliary variable minDis is initialized;

index is initialized to -1;

for i = 0 to k-1 do

{

    dis = distance between instance and i

    If dis < minDis

```
        {
            minDis = dis ;
            index = i ;
        }
    }
    index as key';
    the value of each dimensional coordinate as value';
    output(key', value');
}
```

**3. Reduce function: updating the new center of the coordinates of each cluster, and the new center value of the coordinates written to "the center file".**

```
reduce(<key, value>, <key', value')
{
    initialize an array for storing the cumulative values of the coordinates of
    each     dimension, and the initial value of each component is 0;
    initialize variable num, the total number of samples assigned to the same
    cluster  is recorded, and      the initial value is 0;
    while(value.hasNext())
    {
        from the value.next() analysis the each coordinates and the number num
        of a      sample;
        add the values of each dimension to the corresponding component of the
            array;
        num += num ;
    }
    each component of the array divide by num, and get a new central point
    coordinate;
    key as key';
    constructs a string containing the information of the new central points of
    the       coordinates, the string as value';
```

output(key', value');

}

4. go on until the center of the cluster changes, or the number of iterations reached a preset value.

## 4.4 The algorithm complexity analysis

Parallel computing similarity matrix.Because the subscript i need to calculate the i+n-1 similar values of data points,the time complexity of similar matrix is $O(n+(n-1)+...+1)=O((n^2+n)/2m)$, where 2m is said that the cluster m machine default each machine starts two Map tasks.

Parallel computing the k minimum eigenvectors. Under the condition of non-parallel, time complexity of k different characteristic vectors Laplacian matrix L calculated using Lanczos is $O(kL^{op}+k^2n)$,Where $L^{op}$ is L and vector $V_j$ multiplication. Because we have the matrix L cut into lines stored in the HBase, matrix L and vector Vj multiplied by the distribution of the m machines to run. Under ideal conditions, the time complexity of each multiplication is $L^{op}/m$, the time complexity of the first k eigenvectors parallelization calculated is $O(kL^{op}/m+k^2n)$.

Parallel K-means clustering. Each data point of the new expression $y_i$ is k-dimension, and k center distance calculation in each iteration. Thus, the time complexity of the distance of each data point calculated is $O(k^2)$. So the time complexity of the distance computation for each iteration is $O(nk^2)$. Ideally, all data points are calculated from the distance of the average distribution to the machines, so the time complexity is reduced to $O(\frac{nk^2}{m})\times(num\ of\ iterations)$.

So, the total time complexity should be the sum of the three parts: $O(\frac{n^2+n}{2m})+O(kL^{op}/m+k^2n)+O(\frac{nk^2}{m})\times(num\ of\ iterations)$. It is obvious that the complexity of this time is much smaller than O(n³) and the speed of clustering is improved.

## 4.5 Summary

This chapter briefly introduces the necessity and possibility of parallel spectral clustering. Then, the implementation of the algorithm in Hadoop is described, and the corresponding description is carried out with the pseudo code. In this chapter finally to parallel spectral clustering algorithm the algorithm complexity of the corresponding analysis, in the next chapter will put these algorithms are applied to the actual cluster computing in, to verify the parallel spectral clustering algorithm has special advantages and Hadoop cluster brings speedup.

# Chapter 5 Spectral clustering algorithm results in parallel

The experiment platform hardware configuration: The host 11 (Intel(R) Core(TM) i5-2300 2.80GHz), the memory of 4 GB, operating system (64 - bit Linux Ubuntu). Among them: a Master machine as the name of the node and job control node, only store metadata file blocks and to control the distribution and operation. The other 10 Slave machine as the actual storage nodes and the compute nodes and they are responsible for the storage of file blocks the execution of real data and computing tasks.

## 5.1 Empirical Data

The source of the experimental data represent the topology of a text file. There are two per line to one or more spaces separated string. T is representative figure, v represents a vertex, behind of no. 0 1 representative on the edge of the label is 1. E is for an edge, 0 1 2 represents the connection 0 1 point on the edge of the label is 2. A total of 10029 points and 21054 side. According to the topology of a text file as shown in the figure below:

**Fig.4. The topology of the original text file**

## 5.2 Experiment results

The result of running under different Slave sets. From the results it can be seen in table 5-1, parallel spectral clustering algorithm has been realized, and with the increase of number of the machine, the time needed for parallel computing is less and less. Table 5-1 reflect the parallel computing similarity matrix, parallel computing eigenvalues and eigenvectors, parallel K - means algorithm is accelerated. From the table we can see that with the increase of number of the machine, the three parallel steps have corresponding approximate linear growth. This also proves that under the condition of the Hadoop distributed, spectral clustering has dramatically improved the speed of the algorithm.

But increasing 8 machine more than 10 sets of (as in the figure of 10 machine), algorithm and the steps are not evident on the speed of ascension, in some cases, even slower, this is because for each data set of size n, and the calculation of specific steps, the Hadoop distributed environment have the corresponding critical machine number, within this threshold calculation speed with approximate linear speed is growing. Communication between machine and once is greater than the critical value, the consumption of the growth is even larger than distributed computing, so the speed with the increase of number of the machine would fall.

Table.1. The acceleration of the parallel spectral clustering algorithm based on Hadoop

| Slave Number | Parallel --similarity matrix | Parallel -- k eigenvectors | Parallel--K-means algorithm | Total Time |
|---|---|---|---|---|
| 1 | 1:41:46 | 2:28:14 | 0:28:45 | 4:24:45 |
| 2 | 0:58:45 | 1:45:47 | 0:22:36 | 3:11:08 |
| 4 | 0:30:56 | 1:25:10 | 0:18:09 | 2:28:15 |
| 6 | 0:23:23 | 1:10:44 | 0:14:46 | 1:47:53 |
| 8 | 0:21:15 | 1:00:19 | 0:12:59 | 1:34:33 |
| 10 | 0:22:29 | 1:01:39 | 0:11:45 | 1:35:53 |

As we can see from figure 5-2, the accelerating trend chart. From 1 to 2 sets of that kind of transition, basically is to reduce the time or so commonly. Number of the machine after redouble, speedup growth began to slow, until finally, with the increase of network communication and task allocation overhead, more machines will gradually lose the advantages brought by the machine more.

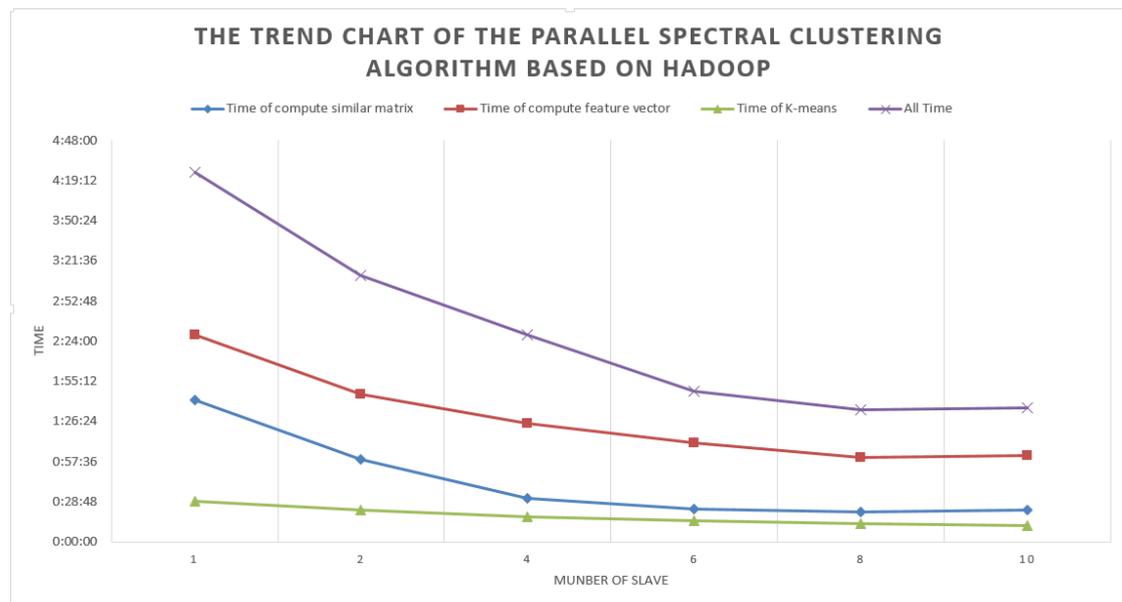

**Fig.5. The trend chart of the parallel spectral clustering algorithm based on Hadoop**

# Bibliography

[1] Lars George. HBase The Definitive Guide[M].CA：O'REILLY 2011

[2] Peng Liu. Hadoop - the shortcut to the cloud computing [M]. Beijing, Electronic Industry Press, 2011-09

[3] E. Gropp and A. Skjellum. Using MPI-2：Advanced Features of the Message -Passing Interface. MIT Press，1999

[4] S. Mahadevan. Fast Spectral Learning Using Lanczos Eigenspace Projections. In AAAI，2008：1472-1475

[5] GAO Yan GU Shi-Wen TANG Jin CAI Zi-Xing. Research of spectral clustering method in machine learning [J]. Computer Science，2007，34(2)：201-203.

[6] FIEDLER M. Algebaric connectivity of graphs [J]. Czechoslovak Mathematical Journal，1973，23(2)：298-305.

[7] Licheng Jiao，Fang Liu，Shuiping Gou. Intelligent Data Mining and Knowledge Discovery [M]. Xi'an: Xidian University Publishing house，2006.

[8] Jain A,Murty M,Flynn P. Data clustering:A Review[J].ACM COMPUTING SURVEYS,1999,(03):264-323.doi:10.1145/331499.331504.

[9] SHI J，MALIK J. Normalized cuts and image segmentation [J]. IEEE Transactions on Pattern Analysis and Machine Intelligence，2000，22(8)：888-905

[10] I.Dhillon，Y.Guan, and B.Kulis.Weighted Graph Cuts without Eigenwectors：A Multilevel Approach. IEEE Trans. On Pattern Analysis and Machine Intelligence，2007，29(11)：1944-1957